\documentclass{article}
\usepackage{spconf,amsmath,graphicx}
\pdfoutput=1 
\usepackage{bbm}
\usepackage{enumitem}

\usepackage{caption}
\usepackage{float}
\usepackage{subcaption}

\title{Motif-based Analysis of Power Grid Robustness under Attacks}
\name{Asim Kumer Dey, Yulia R. Gel, H. Vincent Poor}
\twoauthors
  {Asim Kumer Dey, Yulia R. Gel \sthanks{Yulia R. Gel has been partially supported by NSF IIS 1633331. This material is based in part upon work supported by DARPA. This work was initiated while the second two authors were Visiting Scholars at the Isaac Newton Institute for Mathematical Sciences, Cambridge, UK, and was supported by EPSRC grant no EP/K032208/1. The authors would like to thank Mart\'i Rosas-Casals, for providing the data for European power grid networks and Yongwan Chun for help with GIS data extraction.}}
	{University of Texas at Dallas\\
	Department of Mathematical Sciences\\
	Richardson, TX 75080, USA}
  {H. Vincent Poor}
	{Princeton University\\
	Department of Electrical Engineering\\
	Princeton, NJ 08544, USA}
\begin{document}
%
\maketitle
\begin{abstract}
Network motifs are often called the building blocks of networks.
Analysis of motifs is found to be an indispensable tool for
understanding  \textit{local} network structure, in contrast to measures based on node degree distribution and its functions that primarily address a \textit{global} network topology.
As a result, networks that are similar in terms of global topological properties may differ noticeably at a local level. In the context of power grids, this phenomenon of the impact of local structure has been recently documented in fragility analysis and power system classification. At the same time, most studies of power system networks still tend to focus on global topological measures of power grids, often failing to unveil hidden mechanisms behind vulnerability of real power systems and their dynamic response to malfunctions.  In this paper a pilot study on motif-based analysis of power grid robustness under various types of intentional attacks is presented, with the goal of shedding light on local dynamics and vulnerability of power systems.
\end{abstract}

\section{Introduction}
\label{sec:intro}

The past decade has seen increasing interest in the application of tools developed in the interdisciplinary field of complex network analysis to improve our understanding of power system behavior (for overviews see, e.g.,~\cite{PaganiAiello2013, Cuadra_et_al2015, Rohden_et_al2017} and references therein). Indeed, a power grid can be naturally described as a graph where nodes represent, e.g., transformers, substations or generators, and edges represent electrical connections. Methods of complex network analysis have provided new insights into the fundamental and intrinsic characteristics of power system efficiency, vulnerability and resilience. In particular, numerous recent results indicate that both the topological and functional structure of power grid networks can dramatically impact power system reliability and the effectiveness of associated risk mitigation strategies~\cite{CotillaSanchez_et_al2012, SanchezGarcia_et_al2014, Mureddu_et_al2016}.
Among the most widely explored characteristics of power grid networks are node degree distribution, mean degree, small world properties and, to a lesser extent, betweenness centrality measures -- that is, primarily lower-order connectivity features that are investigated at the level of individual nodes and edges.
However, a number of recent studies of power system reliability indices and stability estimation
suggest that resilience of the power grid is also intrinsically connected to higher-order network features, or network \textit{motifs}~\cite{Menck_et_al2014, Schultz_et_al2014}. The core idea is that if a particular subgraph structure such as, for instance, a triangle, star, square or wheel, occurs significantly often, then such a subgraph likely plays an important role in network functionality. And while higher-order network features have proven to play a fundamental role in understanding organization, functionality and hidden mechanisms behind many complex systems, from brain connectome to  protein-protein interactions to transportation congestion~\cite{milo2002network, prvzulj2007biological, benson2015tensor, ahmed2016kais}, systematic analysis of network motifs in power grids and their impact on system resilience is still largely understudied~\cite{Cuadra_et_al2015, Menck_et_al2014, Schultz_et_al2014, RosasCasals_CorominasMurtra2009} but constitutes an emerging research direction.

In this paper we present a pilot study on motif-based analysis of power grid vulnerability under various types of intentional attacks. In particular, we consider the dynamics of 4-node connected motifs in eight European power grids under three attack strategies, namely, attacked nodes are selected based on degree centrality, betweenness centralities or decreasing order of load (i.e., cascading failures). As a reference, we use a power grid fragility classification of~\cite{RosasCasals_CorominasMurtra2009} based on a tail function of grid degree distribution, that is, deviation of the observed grid cumulative degree distribution from an exponential model. We find that local motif-based properties of fragile and robust networks noticeably differ in terms of their sensitivity to a type of attack. Although a pilot study, these findings suggest that motifs can be useful metrics to characterize a level of power system vulnerability to various types of attacks and certain motifs can potentially serve as early warning indicators of system failure.

\section{Motif-Based Analysis of Power Grids}
\label{sec:motif1}

\textbf{Background on graphs}
We consider an undirected graph $G=(V, E)$ as a model of a power grid network. Here $V$ is a set of nodes, and $E$ is a set of edges. The order and size of $G$ are defined as the number of nodes and edges in $G$, i.e., $|V|$ and $|E|$, respectively. We assume that if an edge $e_{uv} \in E$, then $u\neq v$.
A graph $G$ is connected if there exists a path from any node to any other node in $G$. The distance $d(u,v)$ is defined as the minimum  path length from $u$ to $v$ in $G$. The degree of a node $u$ is the  number of edges incident to $u$. 

 A graph $G'=(V', E')$ is a \textit{subgraph} of $G$ (i.e., $G'\subseteq G$), if $V'\subseteq V$ and $E'\in E$. If $G'=(V', E')$ is a subgraph of $G$ and $E'$ contains all edges $e_{uv}\in E$ such that $u, v\in V'$, then $G'$ is called an \textit{induced} subgraph of $G$.
Two graphs $G'=(V', E')$ and $G^{''}=(V^{''}, E^{''})$ are called \textit{isomorphic}
if there exists a bijection $h: V' \to V^{''}$ such that any two nodes $u$ and $v$ of $G'$ are adjacent in $G'$ if and only if nodes $h(u)$ and $h(v)$
are adjacent in $G^{''}$.

\textbf{Network motifs and their measures} Analysis of higher-order structures of $G$, or multiple-node subgraphs, provides invaluable insights into network functionality and organization beyond the trivial scale of individual vertices and edges.
A \textit{motif} here is broadly defined as a recurrent multi-node subgraph pattern that tends to appear more often than would be expected in a randomized network. Network motifs were introduced by~\cite{milo2002network} in conjunction with the assessment of stability and robustness of biological networks, and later have been studied in a variety of contexts (see \cite{Weihe2015, ahmed2016kais} and references therein). 

\begin{figure}
	\centering
	\includegraphics[width=0.3\textwidth,height=0.15\textheight, bb=198 101 515 328, clip]{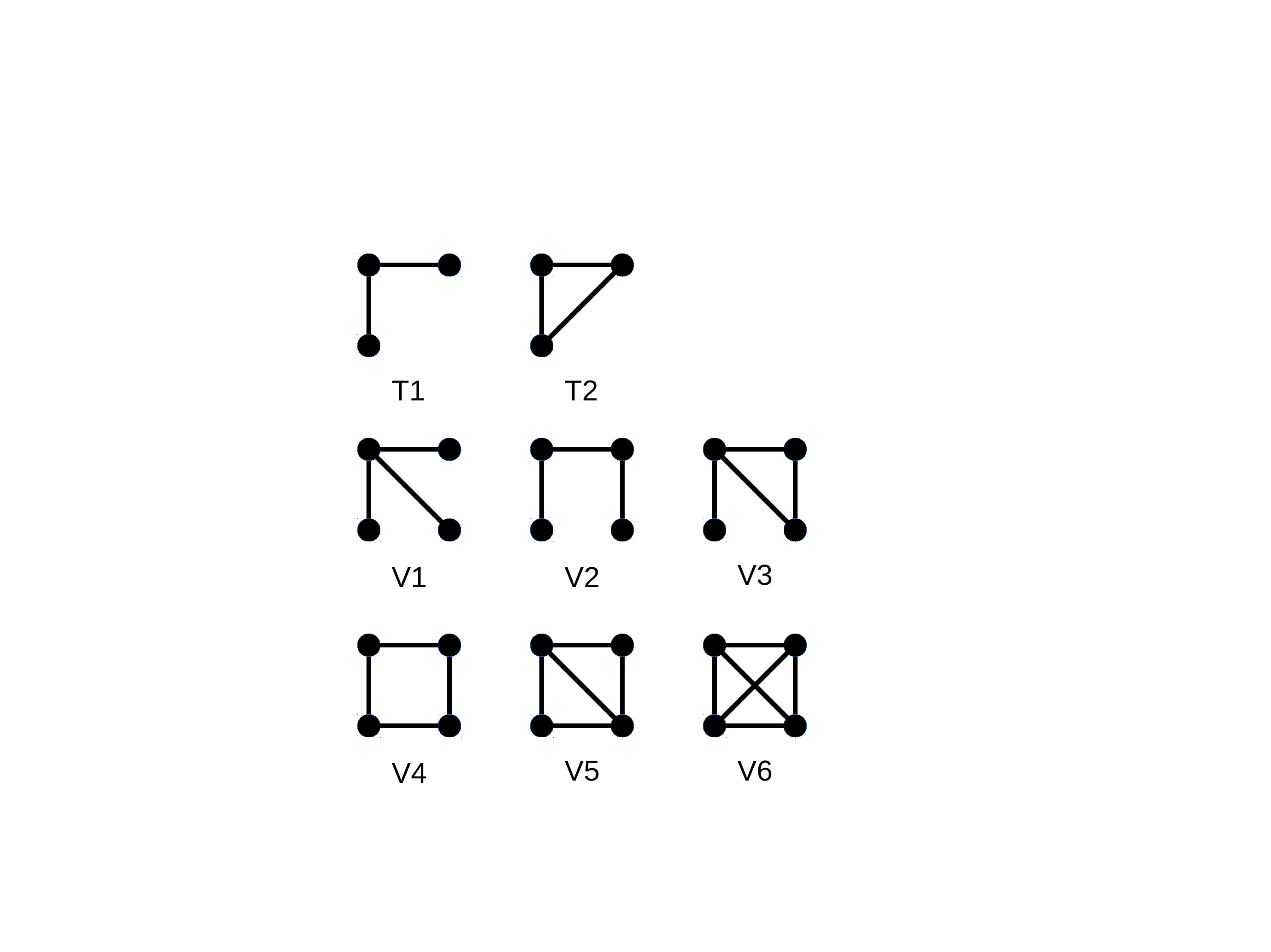}
	\caption{All connected 4-node motifs.}
	\label{F3}
\end{figure}

Formally, let $G_k=(V_k, G_k)$ be a $k$-node subgraph of $G$. If
there exists an isomorphism between $G_k$ and $G'$, $G'\in G$, we say that there exists an \textit{occurrence}, or \textit{embedding} of $G_k$ in $G$. A \textit{motif signature} $f_{G}(G_k)$ is a number of occurrences of $G_k$ in $G$. If a subgraph $G_k$ occurs more frequently than expected by chance, it is called a network  \textit{motif}~\cite{milo2002network}. Figure~\ref{F3} shows all connected 4-node motifs.

Significance of motifs for a particular network can be measured by calculating motifs concentration and $Z$-scores. In $Z$-score the number of appearances of a motif in the observed network is compared with the corresponding quantity for a randomized network: $Z ={{M_R - M}/{s}}$, where $M$ is the mean number of specific motif occurrences in $B$ replicated randomized networks; $s$ is the corresponding sample standard deviation, and $M_R$ is the observed number of motifs in the specific power system network. In this study the randomized networks are simulated using a configuration model, that is, a random graph with the same degree sequence as the observed power grid~\cite{Kolazcyk:2009}. The concentration ($C_i$) of $n$-node motif type $i$ is the ratio between its number of appearances ($N_i$) and total number of $n$-node motifs in the network: $C_i = {N_i / \sum_{i} N_i}$, where $\sum_{i} N_i$ is the total number of $n$-node motifs.

\textbf{Conventional graph characteristics and vulnerability metrics}
The vulnerability of a network can be described as the drop in performance of a network, e.g., power grid network, when a disruptive event emerges. According to~\cite{Cuadra_et_al2015}, the common topology-based vulnerability/robustness metrics are: degree distribution, average path length (APL), diameter (D), clustering coefficient (CC), betweenness centrality (BC), etc.

The node degree of a network is characterized by a probability mass function $P(k)$ indicating the probability that a randomly selected node has $k$ links. As suggested by~\cite{Sole_et_al2008}, higher heterogeneity of power grids and, in particular, higher deviations from the Poisson distribution, tends to imply higher fragility. Power grid networks are assumed to follow exponential cumulated degree distributions~\cite{RosasCasals_CorominasMurtra2009}. That is, the probability that a node chosen uniformly at random has a degree $k$ or higher follows: $P(K \ge k) = C  \exp{(− k\gamma)}$, where $C$ is a normalization constant, $k$ is the node degree and $\gamma$  is a characteristic parameter. According to~\cite{Sole_et_al2008} and~\cite{RosasCasals_CorominasMurtra2009}, a power grid is robust if $\gamma<1.5$ and fragile if $\gamma>1.5$.

\textbf{Robustness under attacks} In robustness under attacks, the aim is to evaluate how a network behaves when a fraction of random or selective nodes are removed. If the node to be removed at each step is selected at random, then the strategy is called a \textit{random attack}. In the case of intentional attacks, the targeted nodes to be removed are selected based on their properties. For instance, if the nodes are selected in the decreasing order of their degree or betweenness centrality,
the resulting attacks is called \textit{degree based attack} or \textit{betweenness based attack}, respectively.
Finally, in a \textit{cascading attack}, nodes are targeted in the decreasing order of their load.
Typically the vulnerability of a network is measured on the basis of the remaining connectivity, largest subgraph size, diameter (D), average shortage path length (APL), etc., after each node removed with different attack strategies. In this study we focus on remaining motif distributions under different targeted attacks, e.g., degree based, betweenness based and cascade attacks. More specifically, our goal is to analyze the decaying rate of a specific motif concentration under different types of attacks and enhance our understanding of local robustness properties of the corresponding network.

\section{Case Studies}
\label{sec:case1}
\begin{figure*}[!ht]
	\centering
	
	\subcaptionbox{Germany}{\includegraphics[width=0.49\textwidth]{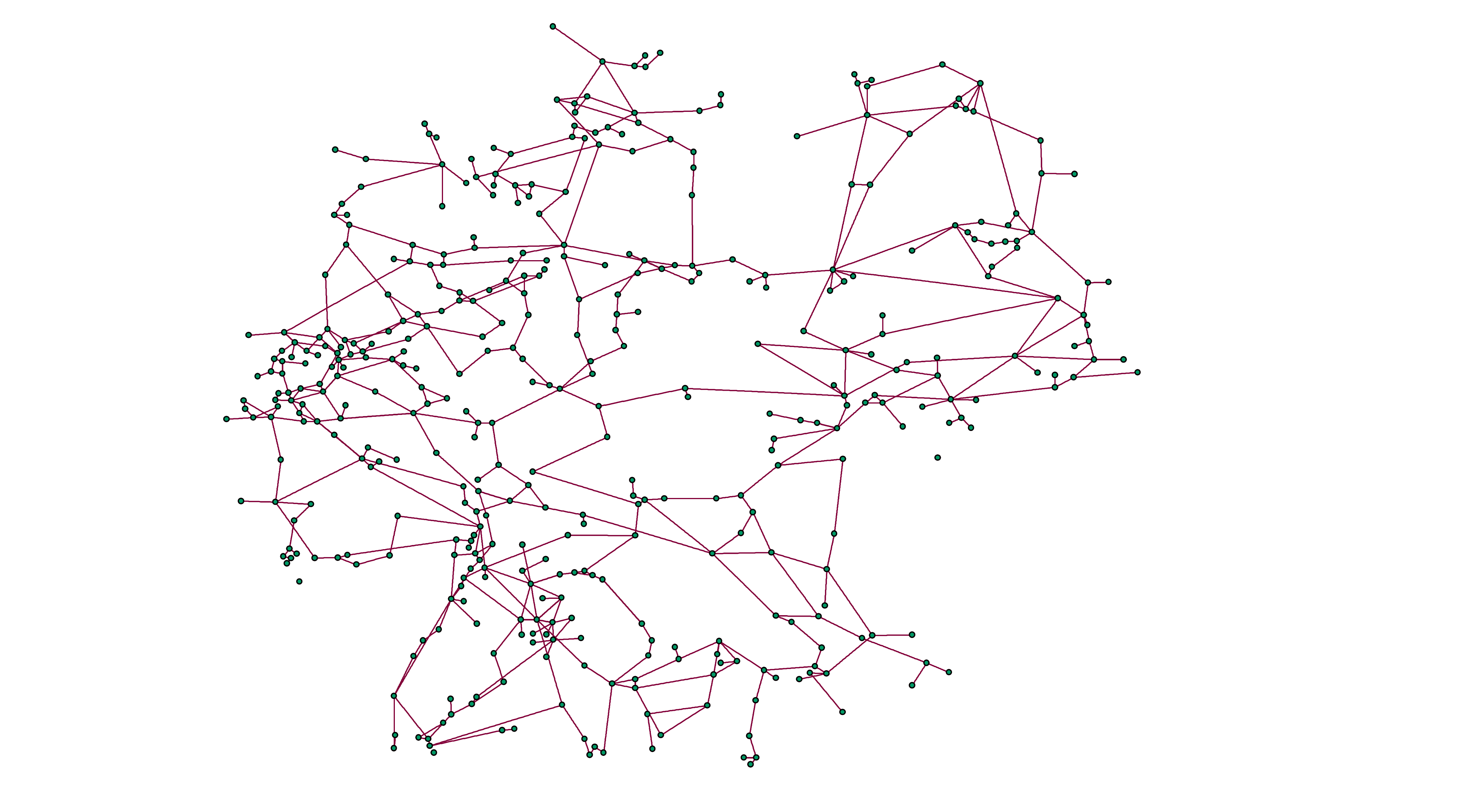}}
	\hfill
	\subcaptionbox{Italy}{\includegraphics[width=0.49\textwidth]{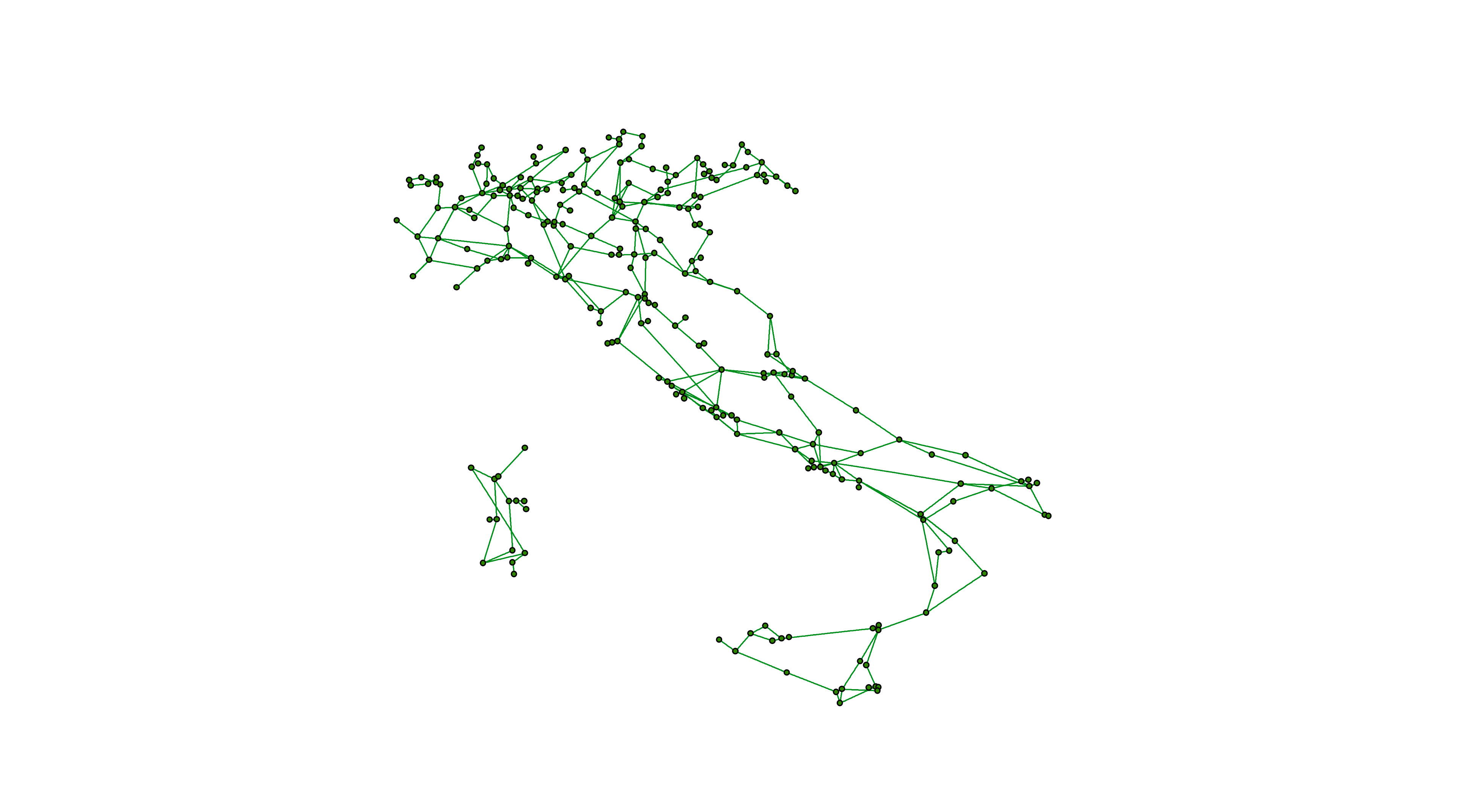}}
	\hfill
	\subcaptionbox{France}{\includegraphics[width=0.49\textwidth]{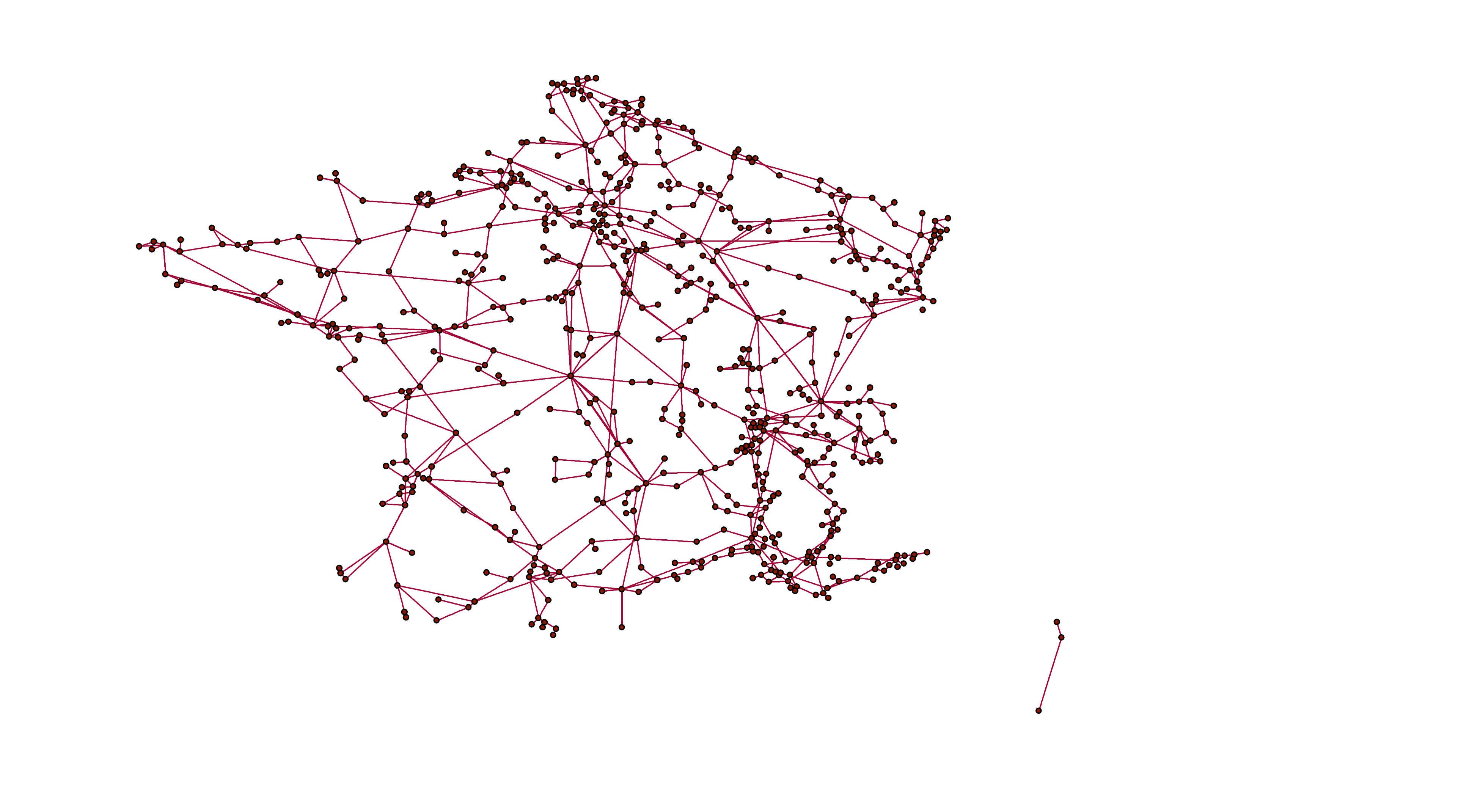}}
	\hfill
	\subcaptionbox{Spain}{\includegraphics[width=0.49\textwidth]{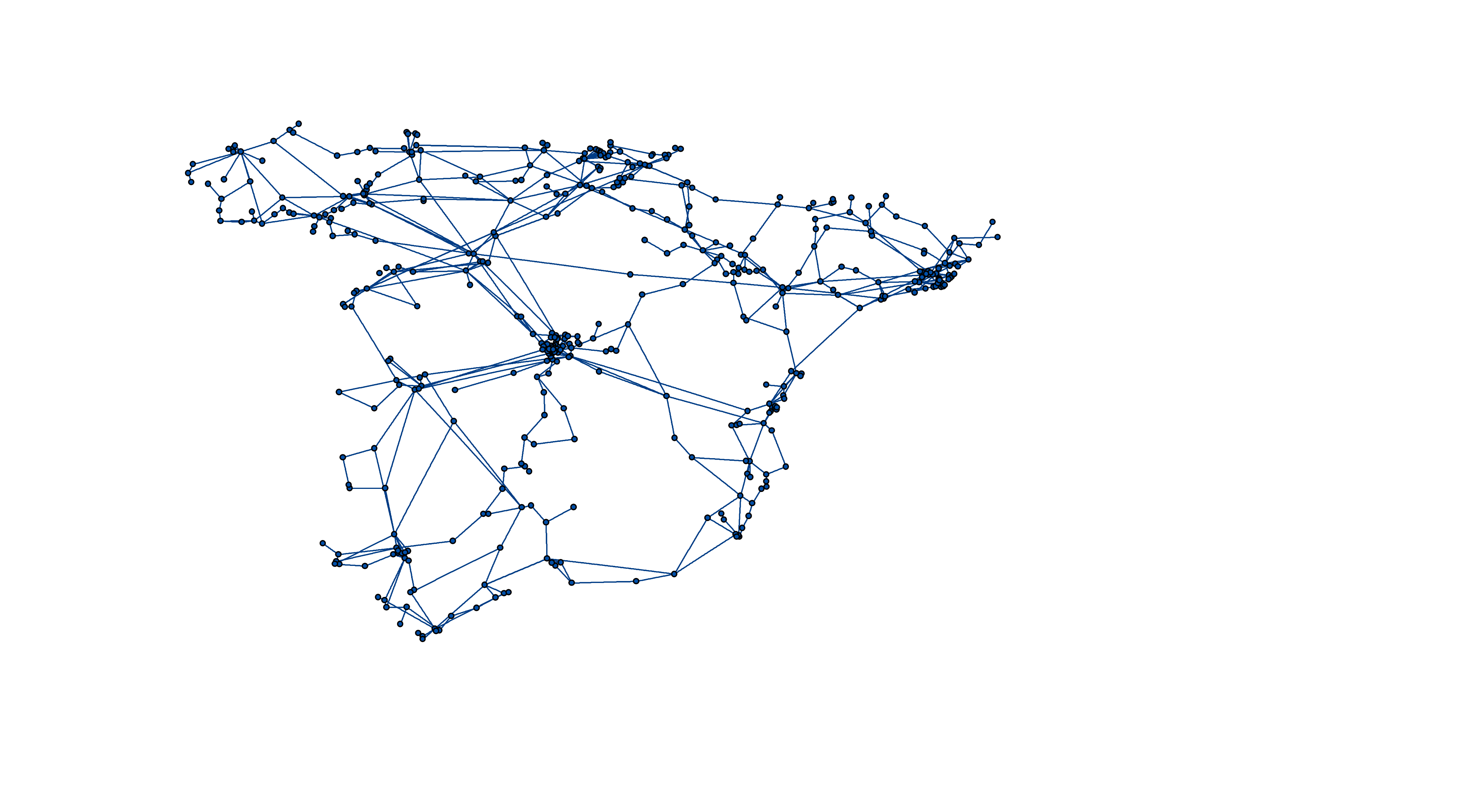}}
	\caption{Maps representing the four European countries power grid networks, where nodes indicate Power stations/sub-stations.}
	\label{Map1}
\end{figure*}

\textbf{Data} In this project we study electricity transmission networks of four European countries, e.g., Germany, Italy, France, and Spain (the same power grid data that are analyzed by~\cite{Luo_RosasCasals2015}), and four European power system operators, e.g., RTE, Amprion, 50 Hertz, and TenneT. The number of nodes e.g., power stations/ sub-stations and edges of the eight power systems are listed in  Table~\ref{T0}. RTE is the French high-voltage transmission system, Amprion is one of the six transmission system operators in Germany, 50Hertz operates in the northern and eastern part of Germany, with a direct connection to Poland, Czech Republic, and the Denmark. TenneT operates in Netherlands and Germany. The network topology of four European countries power systems are shown in Fig.~\ref{Map1}.

\begin{table}[!ht]
	\caption{Network descriptions}
	\label{T0}	
	\begin{center}
		\begin{tabular}{l*{6}{c}r} \hline
			Power  & $\#$ of nodes & 	$\#$ of edges \\
			\hline
			Germany   & 417 & 537 \\		
			Italy & 254 & 357  \\
			
			France & 647 & 880\\
			Spain & 461 & 664 \\
			\hline
			TenneT & 79 & 80\\
			
			RTE    & 190 & 224 \\
			Amprion	& 193 & 252 \\	
			
			50Hertz & 63  & 82 \\
			\hline
		\end{tabular}
	\end{center}
\end{table}

Data for Germany, Italy, France, and Spain are obtained from the Union for the Coordination of the Transmission of Electricity (UCTE) and data for the four system operators are obtained from the SciGRID project~\cite{SciGRID}.

\subsection {Conventional network robustness analysis}
Table~\ref{T1} presents conventional network-based vulnerability metrics for the eight power grids. Table~\ref{T1} suggests that mean degree, average clustering coefficient (CC) and betweennness centrality (BC) for Germany, Italy, and TenneT tend to be lower than the respective metrics for France, Spain, RTE, Amprion, and 50Hertz.
In addition, Table~\ref{T1} shows the estimated fragility parameter $\gamma$, resulting from approximating the cumulative degree distribution of each grid with an exponential model~\cite{Sole_et_al2008, RosasCasals_CorominasMurtra2009}. We find that the electricity transmission systems of Germany and Italy are classified as robust with $\gamma$ of 1.324 and 1.204, respectively.
The TenneT power system is on the border of robustness with $\gamma$ of 1.501. RTE, Amprion and 50Hertz tend to be fragile with $\gamma\approx 2$. France, Spain, and 50Hertz are classified as most fragile grids, with  $\gamma >2$.

\begin{table}[!ht]
	\caption{Vulnerability metrics for the power grid networks}
	\label{T1}	
	\begin{center}
		\begin{tabular}{l l c r ll r} \hline
			Power  & $\gamma$ & 	Mean  & APL & 	D & CC & BC\\
			System & & Degree& & & &  \\
			
			\hline
			Germany   & 1.32 & 2.58	& 11.75	& 30 &	0.07 & 2235.80  \\		
			Italy & 1.21 & 2.81 & 9.74 & 28 & 0.08 & 981.85 \\

			France   & 2.16 & 2.72	& 9.59	& 26 &	0.08 & 2750.01  \\		
			Spain    & 2.22 & 2.89 & 8.26 & 18 & 0.09   & 1670.02\\
			\hline	
			
			TenneT & 1.50 & 2.03 & 5.33 & 12 & 0.10 & 78.71\\
			
			RTE    & 1.86 & 2.36 & 8.05& 20 &  0.17	&	379.91  \\
			Amprion	& 1.98 & 2.61 & 7.01& 	18 & 0.09& 530.10\\	
			50Hertz & 2.13 & 2.60	& 5.15	& 14 & 0.15& 120.60\\
			\hline
		\end{tabular}
	\end{center}
\end{table}

\subsection{Motif based power grid robustness analysis}

\textbf{Motif-based robustness analysis} We start from the concentrations and $Z$-scores for different 4-node motifs that appear in the eight European power grid networks. Since motif distributions are highly skewed, standard $z$- and $t$-quantile are no longer appropriate. Hence, we compare the observed motif $Z$-scores with critical values obtained from parametric bootstrap under a configuration model as a reference. 

\begin{table*}[!ht]
	\caption{Concentration and $Z$-scores (in parenthesis) of 4-node motifs; $\ast$ denotes that a motif is significant in the network with 0.05 level of significance. Number of parametric bootstrap replications is 1,000.}
	\label{T6}	
	\centering
	\begin{tabular}{l*{6}{c}r} \hline
		Power System &	V1 & V2 &	V3 & V4 & V5	\\
		
		\hline
		
		Germany & 0.252  &   0.687  &   0.052  &   0.006  & 0.003 \\
		& (-4.85 $^\ast$) & (-6.98 $^\ast$)& (15.74$^\ast$)&(11.33$^\ast$) &(49.07$^\ast$) \\
		
		Italy & 0.229  &  0.697 & 0.058  & 0.013 & 0.003 \\
		& (-4.36 $^\ast$) & (-9.67$^\ast$)& ( 11.12 $^\ast$)&(14.67$^\ast$) &(19.79$^\ast$) \\
		
		France & 0.292 &  0.642  &   0.059 &  0.006 & 0.002 \\
		& (-3.51 $^\ast$) & (-9.36 $^\ast$)& (24.35 $^\ast$)&(19.58$^\ast$) &(50.89$^\ast$) \\
			
		Spain & 0.346 &  0.570 & 0.075 & 0.005 &  0.003 \\
		& (0.87) & (-12.63$^\ast$)& (19.26$^\ast$)&(11.16$^\ast$) &(34.57$^\ast$) \\
		
			\hline
		
		TenneT & 	0.332 &	0.605 	& 0.060  &	0.003  &	0.000 \\
		& (-5.162$^\ast$) & (-1.873)& (5.179$^\ast$)&(1.491) &(-0.032) \\

		RTE	& 	0.262  & 	0.608 & 0.119  & 0.007 &  0.005\\
		& (-13.321$^\ast$) &(-8.002$^\ast$) & (12.451$^\ast$) &( 3.303$^\ast$)	 & (22.101$^\ast$)\\
		Amprion	& 	 0.413 &	0.504 &	0.074 &	0.007  &	0.002\\
		&(-4.412$^\ast$) & (-6.861$^\ast$)& (4.442$^\ast$) & ( 3.171$^\ast$) &(2.912$^\ast$)\\
		50Hertz	&  0.252 & 0.616 	& 0.123  &	0.007 	& 0.002\\
		& 	 (-3.572$^\ast$)& (-3.690$^\ast$)	& 	 (3.831$^\ast$)& (-0.242)	& 	 ( 0.531)\\
		\hline
	\end{tabular}
	
\end{table*}

Table~\ref{T6} summarize the concentrations and $Z$-scores (in parenthesis) for different motifs that appear in the eight European power grid networks.

In the German, Italian, and French power grids, all 4-node motifs are significant, but in case of the power grid of Spain, star-like motifs $V_1$ are not significant. In cases of TenneT and 50Hertz, we find that both grids deliver non-significant concentrations of detour motifs (i.e., $V_4$ and $V_5$). At the same time, the robust TenneT has also a non-significant concentrations of low connectivity  tree-like \textit{dead end} motifs, $V_2$; while
the fragile 50Hertz has a statistically significant concentration of a tree-like motif $V_2$.

Remarkably, in their studies of the European power grid networks, \cite{RosasCasals_CorominasMurtra2009} find
that power system fragility seems to increase as the elements of the grid
become more interconnected and the number of $\{3, 4\}$-node subgraph patterns
such as stars and triplets, begins to increase.  Independently, based on the analysis of synthetic power grids and a case study of the Northern European power system, \cite{Menck_et_al2014, Schultz_et_al2014} show that an abundance of tree-like dead-end 4-node subgraph patterns leads to a loss of stability and degradation of resilience. More recently,
\cite{Mureddu_et_al2016} who study the impact of removing transmission lines with a high betweenness centrality, suggest that fewer connections imply higher security. Hence, the motif analysis of TenneT and 50Hertz may imply that there exists some balance in representation of low connectivity  tree-like and detour motifs, resulting in a relatively stable system. However, there likely
exists some functional nonlinear interaction among low connectivity and detour motifs and network robustness.


  	

  \subsection{Robustness under attacks}

  In robustness under attack we want to evaluate how a power grid network behave when a fraction of random or selective nodes (e.g., node with highest degree or betweenness) are removed. Here our goal is to measure the remaining connectivity i.e., how large are connected components under random and targeted attacks.

\begin{figure*}
	\centering
	\includegraphics[width=1.0\textwidth,height=0.45\textheight]{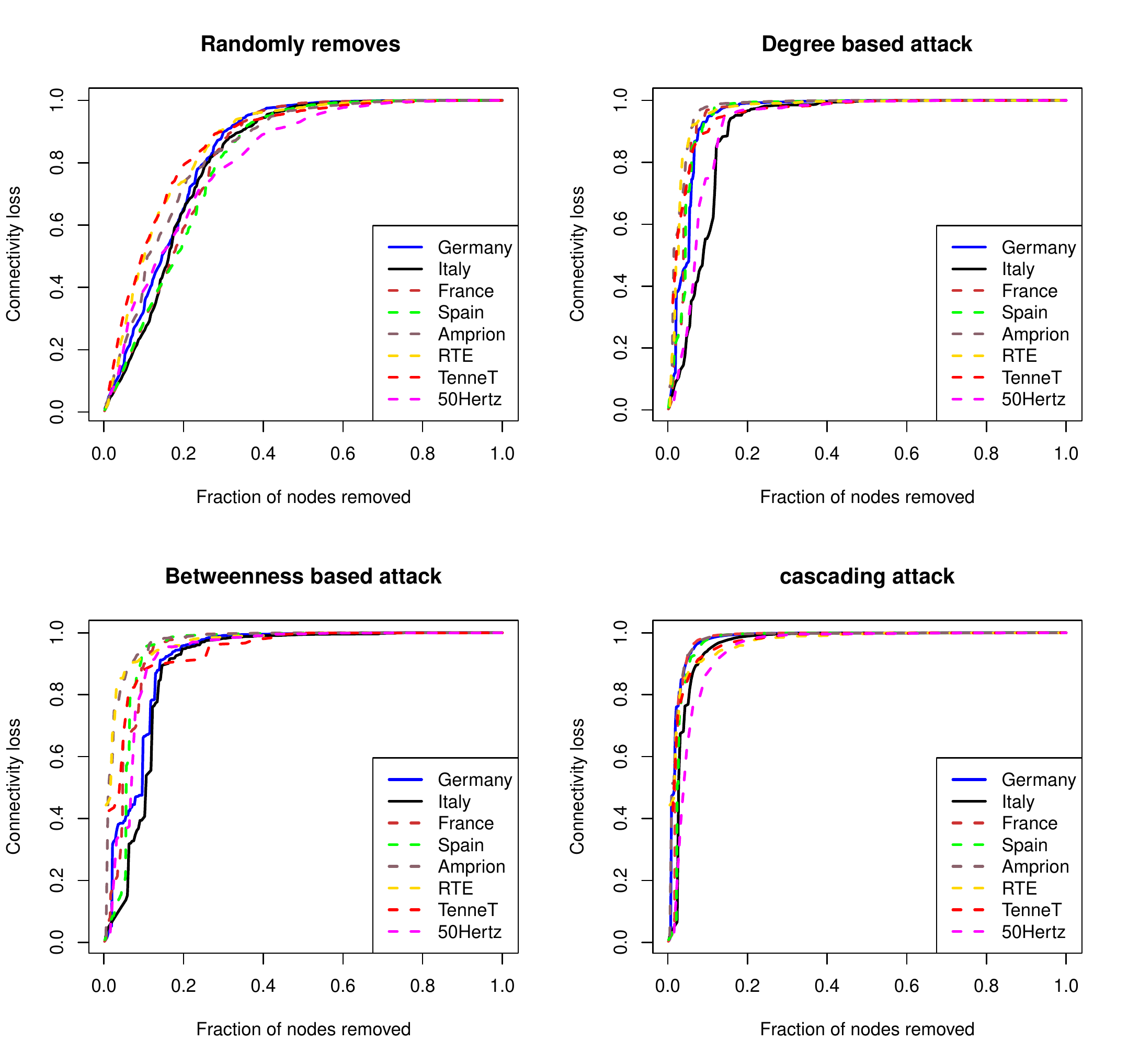}
	\caption {Network robustness under different types of attack.}
	\label{F2}
\end{figure*}

 Figure~\ref{F2} indicates that under cascade, degree-based and betweenness-based attacks, connectivity in the power grid networks of
 France, Spain, RTE, Amprion and 50Hertz
   decays more rapidly than connectivity in the power grid systems of Germany, Italy, and TenneT. However, under random attacks all eight power systems loss connectivity with similar rates.

\begin{figure*}[]
	\centering
	\includegraphics[width=1.00\textwidth,height=0.85\textheight]{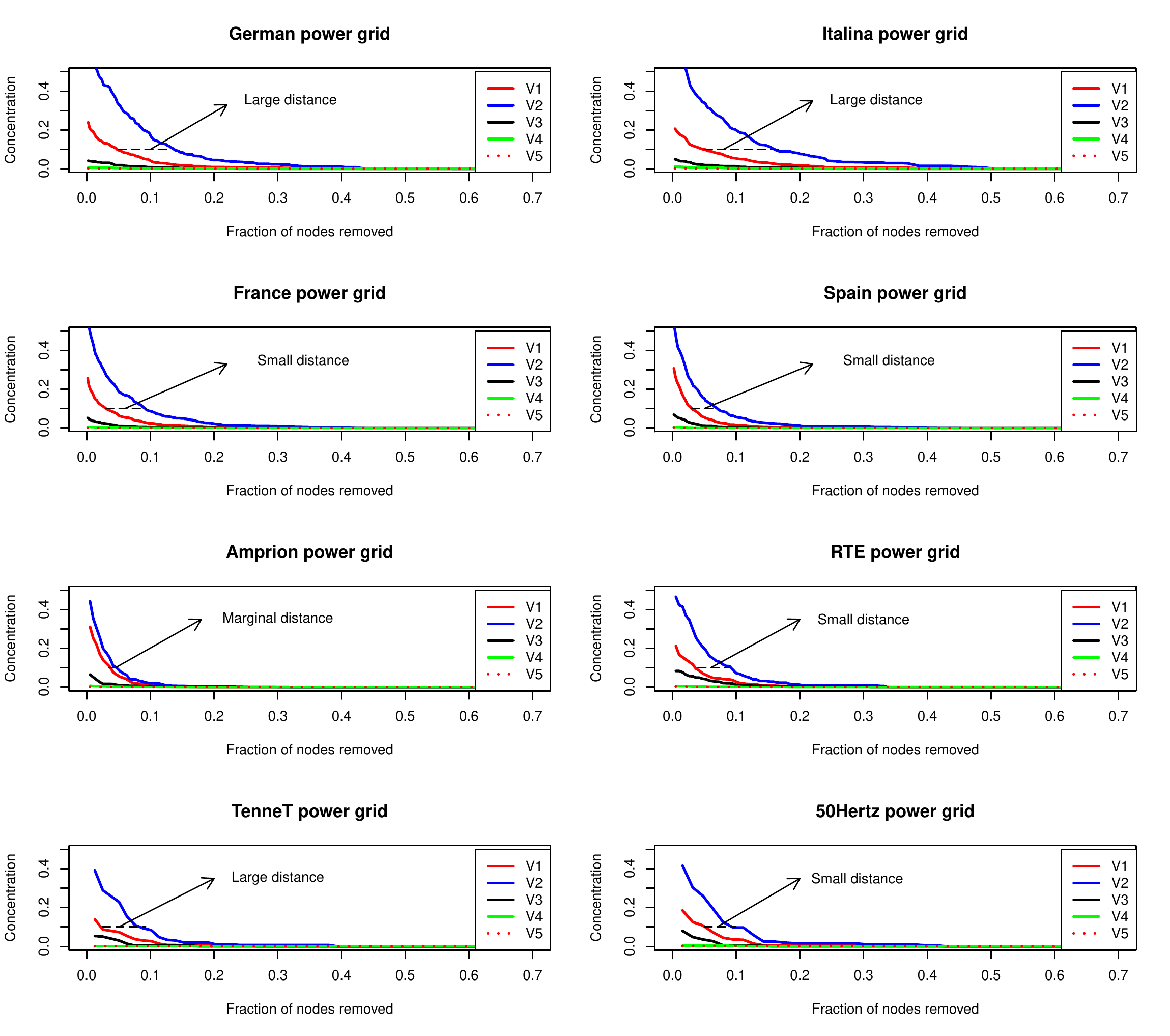}
	\caption {Dynamics of 4-motif concentration under degree based attack.}
	\label{F4}
\end{figure*}

\begin{figure*}[]
	\centering
	\includegraphics[width=1.0\textwidth,height=0.85\textheight]{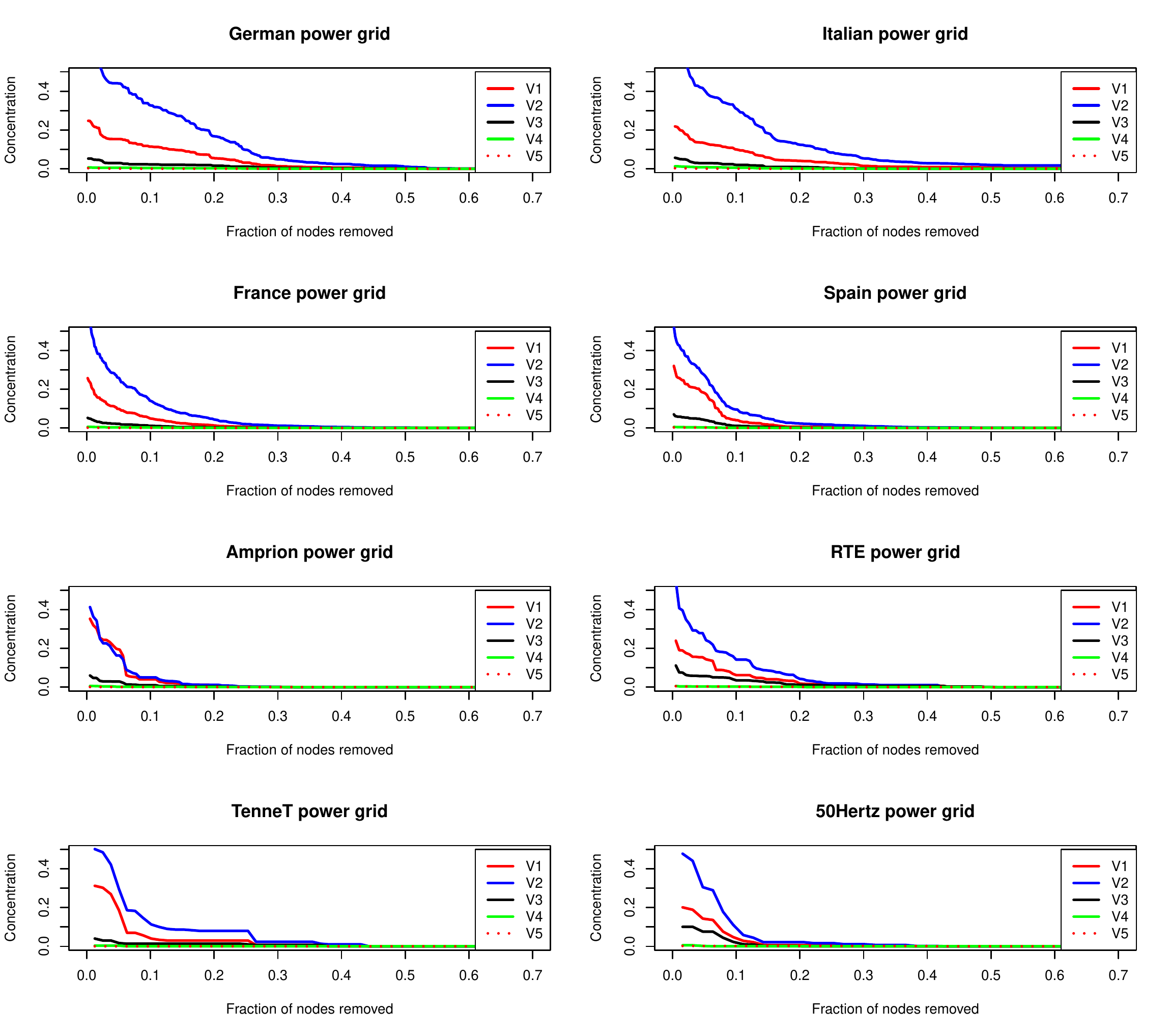}
	\caption {Dynamics of 4-motif concentrations under betweenness based attacks.}
	\label{F5b}
\end{figure*}

\begin{figure*}[]
	\centering
	\includegraphics[width=1.0\textwidth,height=0.85\textheight]{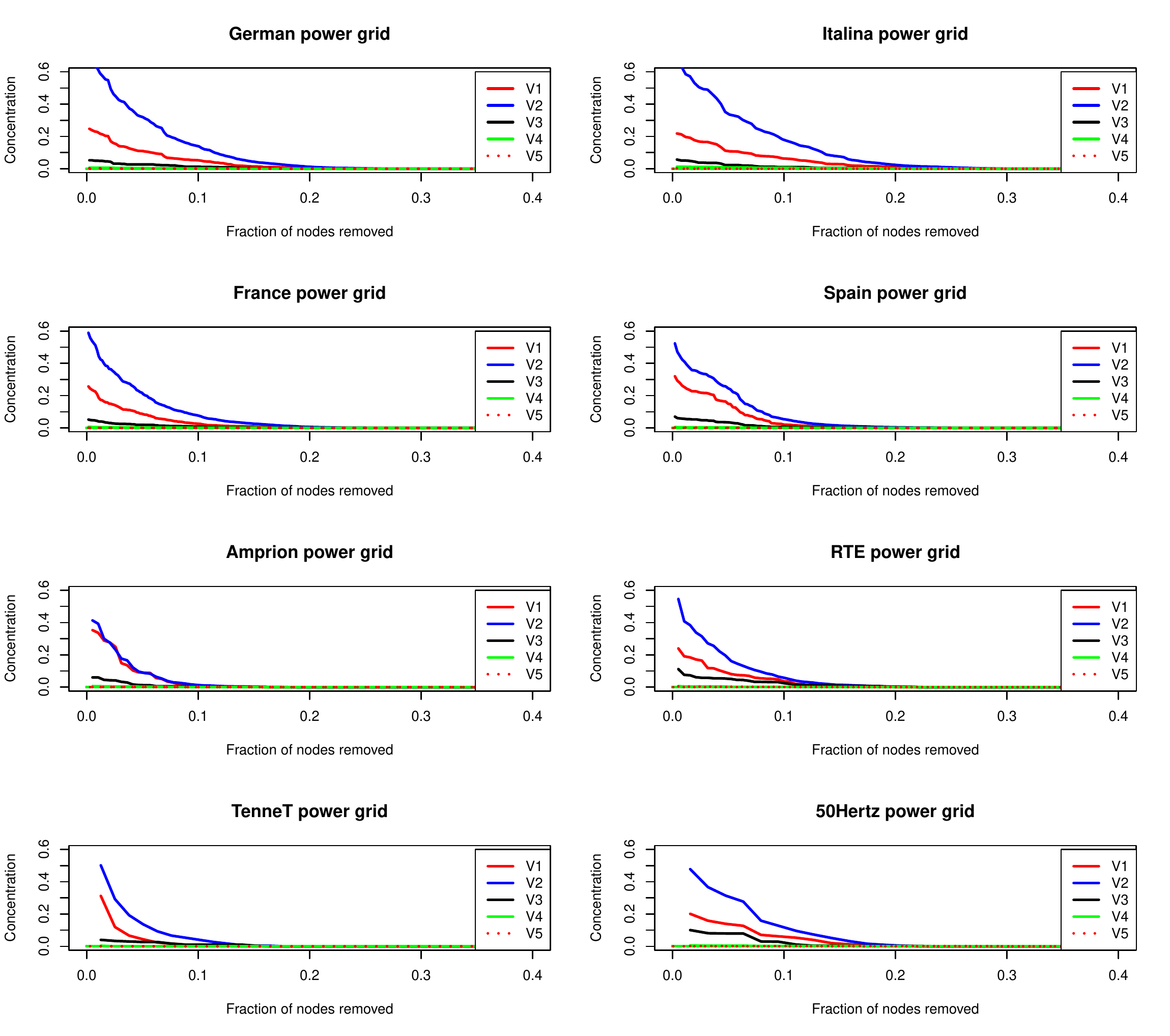}
	\caption {Dynamics of 4-motif concentrations under cascade attacks.}
	\label{F5C}
\end{figure*}

  To assess vulnerability of the eight power grid networks, we also investigate the dynamics of motif distributions under different targeted attacks. Fig.~\ref{F4}--\ref{F5C} depict the decaying rate of different 4-node motif concentrations under three types of targeted attacks, e.g., degree-based, betweenness-based and cascade attacks. The motif concentrations in the networks of France, Spain, RTE, Amprion, and 50Hertz tend to exhibit different dynamics of decay, than the motif concentrations in the power grid networks of Germany, Italy, and TenneT.

  For instance, the decay of motif concentration curves for $V_2$ is noticeably steeper than the decay of motif concentration curves for $V_1$ in the German, Italian, and TenneT power grid networks under degree- and betweenness-based attacks (Fig.~\ref{F4}-\ref{F5b}), whereas, the respective gaps among concentration curves for $V_1$ and $V_2$ in other five networks tend to narrower.
Note that according to~\cite{Sole_et_al2008} and~\cite{RosasCasals_CorominasMurtra2009}, the power grids of Germany and Italy are classified as robust, while the grids of France, Spain, RTE, Amprion, and 50Hertz are classified as fragile. TenneT power grid is a borderline case (see Table~\ref{T1}). 

\begin{figure*}[]
	\centering
	\includegraphics[width=1.0\textwidth,height=0.85\textheight]{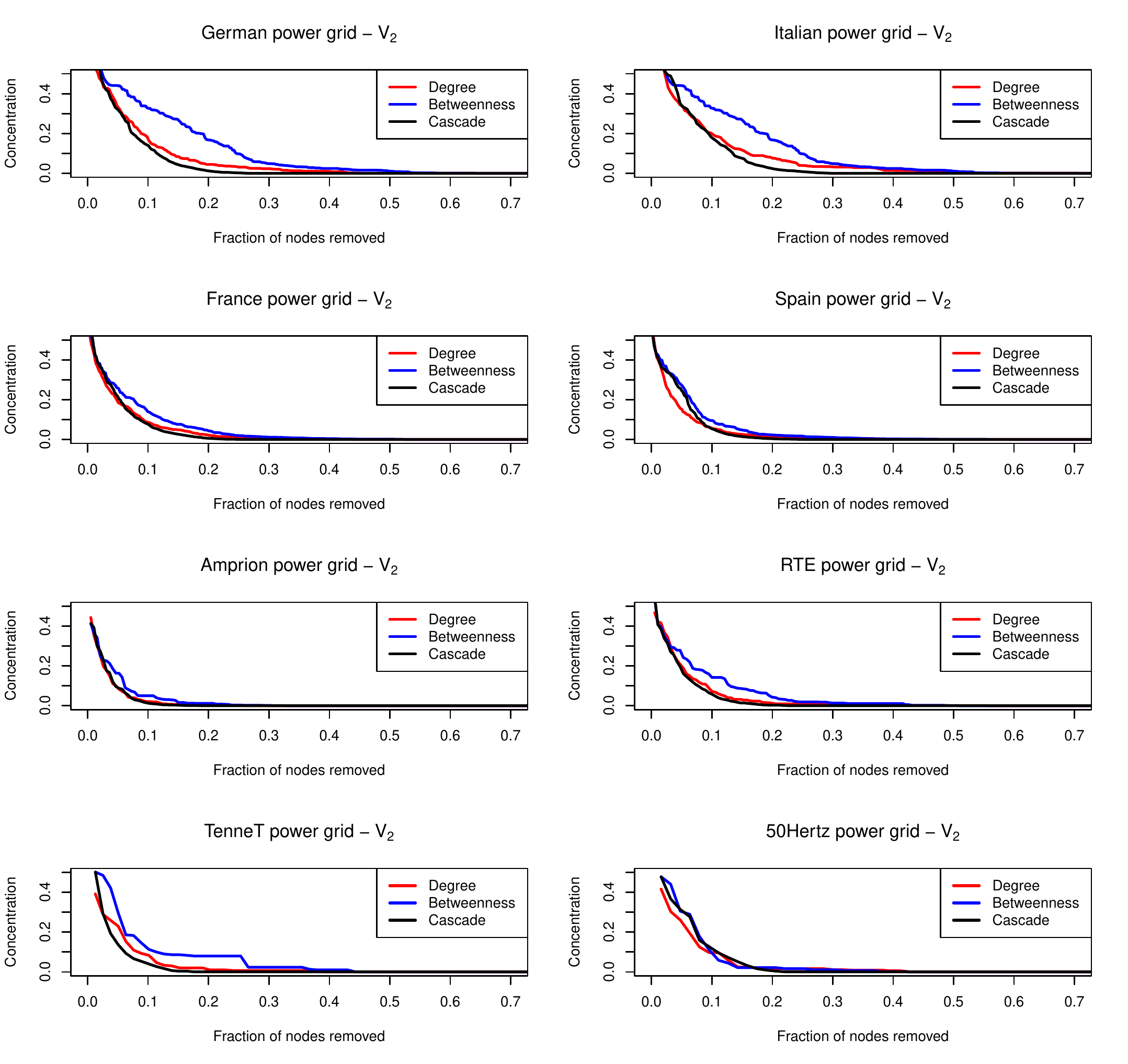}
	\caption {Persistence of $V_2$ motif concentrations under different targeted attacks.}
	\label{F6}
\end{figure*}

\begin{figure*}[]
	\centering
	\includegraphics[width=1.0\textwidth,height=0.85\textheight]{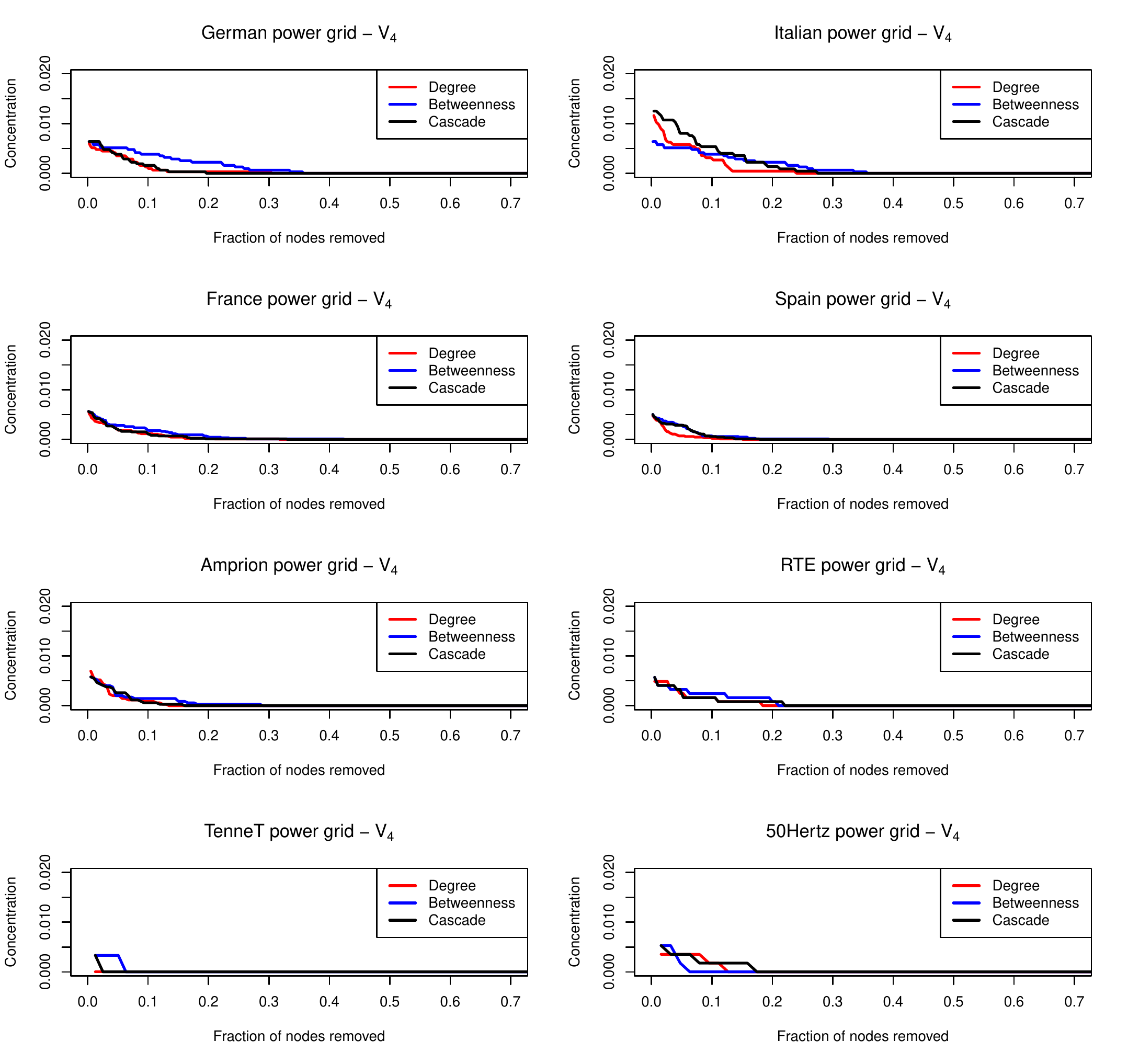}
	\caption {Persistence of $V_4$ motif concentrations under different targeted attacks.}
	\label{F7}
\end{figure*}

 Furthermore, Fig.~\ref{F6} suggest that concentrations of $V_2$ for the power systems of Germany, Italy, and TenneT under betweenness based attacks decay noticeably slower than
  the respective concentrations under degree-based and cascade attacks. However, dynamics of
  concentrations of five other networks appear not to depend on a type of attack.  Similar patterns are observed for concentration of $V_4$, although the difference between concentration curves
    under betweenness-based attacks and under the degree-based and cascade attacks, is less profound than for the case of $V_2$
    (see Fig.~\ref{F7}). (Notice, concentration of $V_4$ for 50Hertz exhibits a steeper decay under betweenness-based attacks than under degree-based and cascade attacks, which is contrast to other systems.) These results suggest that a local motif structures of fragile and robust networks appear to be sensitive with respect to an attack strategy and a considered motif.

\section{Conclusion and discussion}

Although even basic $\{3, 4\}$-node motifs have been proven to unravel hidden mechanisms behind functionality and stability of various complex systems (see~\cite{milo2002network, Weihe2015, Tsourakakis_et_al2016} and reference therein), including a limited number of motif-based vulnerability studies in power networks~\cite{Schultz_et_al2014, RosasCasals_CorominasMurtra2009}, to our knowledge, there exists no previous study of motif-based analysis of power systems under attacks. In this pilot study we focus on motif-based analysis of local power grid vulnerability under random and intentional attacks.
We find that the dynamics of distributions of 4-node motifs under various attacks differ with respect to the global tail-based grid classification of power grid fragility proposed in~\cite{RosasCasals_CorominasMurtra2009}. In particular, we find that robust and fragile
power systems exhibit different degrees of local sensitivity and degradation with respect to the type of attack and the type of motif.
Hence, motif characteristics such motif concentrations can be potentially used as alternative local metrics of fragility under attacks as well as early warning indicators of system degradation and failure.
In the future, we plan to further expand this study into a hybrid analysis of local motif-based topological and functional properties of weighted power grid systems.



\end{document}